\documentclass{article}

\PassOptionsToPackage{square, numbers, sort&compress}{natbib}

\usepackage[preprint]{neurips_2021}

\usepackage[utf8]{inputenc} 
\usepackage[T1]{fontenc}    
\usepackage{hyperref}       
\usepackage{url}            
\usepackage{booktabs}       
\usepackage{amsfonts}       
\usepackage{nicefrac}       
\usepackage{microtype}      
\usepackage{xcolor}         

\usepackage{mathtools,latexsym, enumerate, amsmath, amsthm, amssymb,hyperref,tikz,framed,color,enumitem,enumerate}
\usepackage{graphicx}

\title{Molecular Energy Learning Using Alternative Blackbox Matrix-Matrix Multiplication Algorithm for Exact Gaussian Process}

\author{%
  Jiace Sun \\
  Division of Chemistry and Chemical Engineering\\
  California Institute of Technology\\
  Pasadena, CA 91125, USA\\
  \texttt{jsun3@caltech.edu}\\
   \And Lixue Cheng \\
   Division of Chemistry and Chemical Engineering\\
   California Institute of Technology\\
   Pasadena, CA 91125, USA\\
   \texttt{lcheng2@caltech.edu}\\
   \And Thomas F. Miller III \\
   Division of Chemistry and Chemical Engineering\\
   California Institute of Technology\\
   Pasadena, CA 91125, USA\\
   \texttt{tfm@caltech.edu}
}

\begin{document}

\maketitle

\begin{abstract}

We present an application of the blackbox matrix-matrix multiplication (BBMM) algorithm to scale up the Gaussian Process (GP) training of molecular energies in the molecular-orbital based machine learning (MOB-ML) framework.
An alternative implementation of BBMM (AltBBMM) is also proposed to train more efficiently (over four-fold speedup) with the same accuracy and transferability as the original BBMM implementation.
The training of MOB-ML was limited to 220 molecules, and BBMM and AltBBMM scale the training of MOB-ML up by over 30 times to 6500 molecules (more than a million pair energies).
The accuracy and transferability of both algorithms are examined on the benchmark datasets of organic molecules with 7 and 13 heavy atoms.
These lower-scaling implementations of the GP preserve the state-of-the-art learning efficiency in the low-data regime while extending it to the large-data regime with better accuracy than other available machine learning works on molecular energies.

\end{abstract}

\section{Introduction}
Machine-learning (ML) for quantum chemistry has emerged as a versatile method in chemical sciences in recent years, facilitating innovation in a variety of domains such as molecular modeling~\cite{christensen2020fchl,schutt2018schnet,qiao2020orbnet}, drug discovery~\cite{kwan2018concerted,dirocco2017multifunctional,rosales2019rapid}, and material design~\cite{guerin2018control,hongo2010failure,maurer2019advances,rossi2016anharmonic} by delivering accurate predictions at a low computational cost.
One successful example of such methods is the recently developed molecular-orbital-based machine learning (MOB-ML)~\cite{Welborn2018,Cheng2019,husch2021improved}, which can predict molecular energies with a high degree of accuracy while requiring relatively low amounts of training data.
Nevertheless, extending MOB-ML to the large data regime has remained challenging due to the steep $O(N^3)$ complexity scaling associated with the use of Gaussian Process (GP) regression.
A strategy to reduce the complexity of GP is to introduce a low-rank kernel approximation, which has been exploited in the Sparse Gaussian Process Regression~\cite{sgpr} and Stochastic Variational Gaussian Processes~\cite{svgp} methods.
However, such treatments of GP sometimes result in significant loss of accuracy.
In contrast, Gardner et al.~\cite{gardner2018gpytorch,wang2019exact} recently proposed the blackbox matrix-matrix multiplication (BBMM) method, which provides exact GP inference while reducing the training time complexity to $O(N^2)$ and allowing for multi-GPU usage.

In this work, we employ BBMM and a novel alternative implementation for BBMM (AltBBMM) to speedup and scale the GP training in MOB-ML for molecular energies. We show that AltBBMM delivers more efficient training on over 1 million pair energies without sacrificing transferability across chemical systems of different molecular sizes.
The accuracy and efficiency of BBMM and AltBBMM in modeling physical problems are demonstrated by comparisons with literature results on the same datasets.

\section{Background}
\subsection{Molecular orbital based machine learning (MOB-ML)}
The total energy of a given chemical system can be written as the sum of the Hartree-Fock and correlation energies. The correlation energy, $E_{\text{corr}}$, can be further decomposed into pair energies, $\epsilon_{ij}$, associated with occupied molecular orbitals (MO) $i$ and $j$, such that $E_{\text{corr}} = \sum_{ij} \epsilon_{ij}$.
MOB-ML learns pair energies by 
$\epsilon_{ij}\approx \epsilon^{\text{ML}}(\boldsymbol{f}_{ij})$, where the features $\boldsymbol{f}_{ij}$ describe the interactions between the molecular orbitals.\cite{husch2021improved}
Due to the different scales of the values, the diagonal pair energies $\epsilon_{\text{d}} = \{\epsilon_{ij} \big| i=j\}$ and the offdiagonal pair energies $\epsilon_{\text{o}} = \{\epsilon_{ij} \big| i\neq j\}$ are trained separately, effectively producing two different models.

\subsection{Gaussian Processes (GP)}
Gaussian Processes (GP) are non-parametric kernel-based machine learning methods that predict the probabilistic distribution of unobserved data.
Given the observed data $(X, y), X \in \mathbb{R}^{N\times d}, y \in \mathbb{R}^{N}$ with a Gaussian noise $\sigma^2 \in \mathbb{R}$ and a prior covariance function or kernel $K: \mathbb{R}^{d} \times \mathbb{R}^{d} \rightarrow \mathbb{R}$, the prediction $f(X^{\prime})$ test points $X^{\prime} \in \mathbb{R}^{M\times d}$ is a joint Gaussian distribution such that
\begin{equation}
    \label{eq:pred}
	\mathop{\mathbb{E}}[f(X^{\prime})] = K(X^{\prime}, X) \hat{K} ^ {-1} y, \qquad \text{Var}[f(X^{\prime})] = K(X^{\prime}, X) \hat{K} ^ {-1} K(X, X^{\prime}),
\end{equation}
where $\hat{K} = K(X, X) + \sigma^2 I$.
The hyperparameters,
which include the Gaussian noise and kernel parameters, are learned in GP training by maximizing the log marginal likelihood
\begin{equation} \label{ll}
	L = -\frac{1}{2} y^T \hat{K}^{-1} y - \frac{1}{2} \text{log}|\hat{K}| - \frac{N}{2} \log 2\pi.
\end{equation}
The typical way to calculate the above quantities is the Cholesky decomposition, which has a time complexity of $O(N^3)$ and a memory complexity of $O(N^2)$.
Such complexities limit the training size of GP-based models, such as MOB-ML, to around 50000 data points.

\section{Method}

\subsection{Conjugate gradient (CG)}
The conjugate gradient (CG)~\cite{van2003iterative} algorithm offers another way to obtain the predictive mean in ~Eq.\ref{eq:pred} by iteratively solving $\omega = \hat{K}^{-1} y$, or equivalently $\hat{K} \omega = y$ with an $O(N^2)$ cost in each iteration.
CG requires only the matrix-vector multiplications (MVMs) with the kernel matrix $\hat{K}$, which is amenable to multi-GPU acceleration.
In the iteration $k$, the solution is found in the order-$k$ Krylov space
\begin{equation}
	\mathcal{K}=\operatorname{span}\left\{\hat{K}^{i} y \Big| i=0, 1, ... k-1\right\}.
\end{equation}

The solution $\omega^k$ of CG at iteration $k$ converges to the exact solution $\omega^*$ exponentially measured by the relative residual $\frac{||\hat{K}\omega^k-y||}{||y||}$.
However, the total number of iterations $k_c$ to converge is usually very large for a kernel $\hat{K}$ with high singularity.
A common way to reduce $k_c$ is to construct a preconditioner $P$ and then solve the equivalent equation $P^{-1}\hat{K}\omega = P^{-1}y$ such that $P^{-1}\hat{K}$ is less singular than $\hat{K}$.~\cite{cutajar2016preconditioning}.

Block conjugate gradient (BCG)~\cite{bcg1}, as a variant of CG, can also be used to further reduce $k_c$. It extends CG to solve $s$ linear equations $\hat{K}\omega_{i}=y_{i}, i=0, 1 ... s-1$ simultaneously. The number of linear equations $s$ is also known as block size.
In the iteration $k$ of BCG, the solution is found in
\begin{equation}
	\mathcal{K}^{\text{block}}=\operatorname{span}\left\{\hat{K}^{j} y_i \Big| i=0, 1, ... s-1, j=0, 1, ... k-1\right\}.
\end{equation}
By setting $y_0=y$, and $y_i\sim N(0,I)$ for $i>0$, BCG can converge to the same exact solution $\omega_0^*=\omega^*$ with fewer iterations since $\mathcal{K}_{k} \subset \mathcal{K}^{\text{block}}_{k}$.

\subsection{Blackbox matrix-matrix multiplication (BBMM) and Alternative BBMM (AltBBMM)}
BBMM~\cite{gardner2018gpytorch,wang2019exact} calculates the GP inference by utilizing CG combined with the pivoted Cholesky decomposition preconditioner~\cite{bach2013sharp, harbrecht2012low}.
Furthermore, a modified batched version of conjugate gradients (mBCG\footnote{mBCG (modified batched conjugate gradients) differs from BCG (block conjugate gradient)})~\cite{gardner2018gpytorch} is also proposed to estimate the marginal likelihood and its derivatives, which are required in the GP hyperparameter optimization.
These enhancements reduce the training complexity to $O(N^2)$ in time, and $O(N)$ in memory and therefore enable the training of a million data points.

In this work, we propose an alternative realization of BBMM (AltBBMM) to achieve similar accuracy as BBMM with a lower cost in molecular energy prediction applications, where a low Gaussian noise ($10^{-5}{\sim}10^{-8}$) is required to reach the desired accuracy.
However, sine the low Gaussian noise significantly increases the singularity of $\hat{K}$, CG would converge slowly or even fail to converge when the rounding errors exceed the Gaussian noise.\cite{greenbaum1997iterative}
In order to speedup the CG convergence, we employ the BCG algorithm described in the previous section.
The additional computational cost of BCG in each iteration is negligible compared with the kernel matrix calculations.
To further improve the robustness of the convergence, we use the double-precision floating numbers in the implementation and employ the symmetric preconditioning $P^{-1/2} \hat{K} P^{-1/2}$.
The Nystreoem preconditioner~\cite{cutajar2016preconditioning} is used as an example, but we note that better preconditioners could exist. Finally, the hyperparameters are optimized on a random subset of the entire training set in AltBBMM since the optimized hyperparameters remain similar across various training sizes for MOB-ML. 

\section{Computational details}
We train all the models on random subsets of the QM7b-T dataset~\cite{Welborn2018, Cheng2019,husch2021improved}, which contains 7211 organic molecules with up to 7 heavy atoms. 
The test sets are the remaining QM7b-T molecules and the whole GDB-13-T dataset~\cite{Welborn2018, Cheng2019,husch2021improved} containing 1000 organic molecules with 13 heavy atoms.
The Mat\'ern 5/2 kernel is used in all the GP trainings.
We independently implement the BBMM according to the description of the mBCG and hyperparameter optimization in Ref.~\citenum{gardner2018gpytorch}.
The symmetric Nystroem preconditioner and the block CG are used in this work.
In both BBMM and AltBBMM, the rank $r$ of the preconditioner is chosen as 10000, the BCG block size $s$ is fixed as $50$, and the BCG iterations stop when all the $s$ relative residuals are smaller than $10^{-6}$.
The hyperparameters are optimized from a full GP trained on 50 random molecules.
To overcome the memory limit and maximize the multi-GPU efficiencies, the kernel computations in CG are performed in $4096\times 4096$ batches, and such computations are dynamically distributed to all the available GPUs.
Additionally, we add a Gaussian noise regularization $\sigma^2_{\text{add}} = 10^{-5}$ to the optimized Gaussian noise reduce the singularity of $\hat{K}$.

\section{Results}
\subsection{Low noise regularization for accurate GP}
We first demonstrate the necessity of utilizing a low noise regularization to achieve accurate predictions.
We train all the offdiagonal energies ($\epsilon_{\text{o}}$) pairs from 1000 QM7b-T molecules with different $\sigma^2_{\text{add}}$ and test on the $\epsilon_{\text{o}}$ of the rest QM7b-T molecules.
The training time and the prediction mean absolute error (MAE) are displayed in Table \ref{table:noise}.
For both BBMM and AltBBMM, regularizing with $\sigma^2_{\text{add}} = 10^{-1}$  results in a less singular $\hat{K}$ and saves half the training time, but its prediction MAE doubles when compared to the results of $\sigma^2_{\text{add}} = 10^{-5}$.
Since the MOB-ML data generation is significantly more expensive than model training, we fix $\sigma^2_{\text{add}} = 10^{-5}$ for all of the following BBMM and AltBBMM experiments to achieve the most accurate model with the least amount of data. 

\begin{table}[bhtp]
	\caption{Test MAEs (kcal/mol) of offdiagonal contributions ($\sum \epsilon_\mathrm{o}$) in each molecule and training time (s) by training on  $\epsilon_{\text{o}}$ pairs from 1000 QM7b-T molecules (N=175,795) with different Gaussian noise regularizations.}
	\label{table:noise}
	\centering
	\begin{tabular}{ccccc}
		\toprule
		& \multicolumn{2}{c}{BBMM} & \multicolumn{2}{c}{AltBBMM} \\
		\cmidrule(r){2-5}
		 $\sigma^2_{\text{add}}$ & Test MAE & Time & Test MAE & Time \\
		\midrule
		$10^{-1}$ & 0.636 & 1104.30 & 0.619 & 456.46   \\
		$10^{-5}$ & 0.314 & 2150.93 & 0.312 & 760.54  \\
		\bottomrule
	\end{tabular}
\end{table}

\begin{table}[bhtp]
	\caption{Test MAEs (kcal/mol) and training time (hrs) of BBMM and AltBBMM trained on 6500 QM7b-T molecules$^a$ with the same initial hyperparameters.}
	\label{table:6500}
	\centering
	\begin{tabular}{cccc}
		\toprule
		Algorithm & QM7b-T MAE & GDB-13-T MAE/7HA & Time \\
		\midrule
		BBMM & 0.185 & 0.490 & 26.52 \\
		AltBBMM & 0.193 & 0.493 & 6.24\\
		\bottomrule
		\multicolumn{4}{c}{$^a$ Training size of $\epsilon_o$ is 1,152,157 and training size of $\epsilon_d$ is 124,973}
	\end{tabular}
\end{table}

\begin{figure}[bthp]
    \centering
	\includegraphics[width=0.98\textwidth]{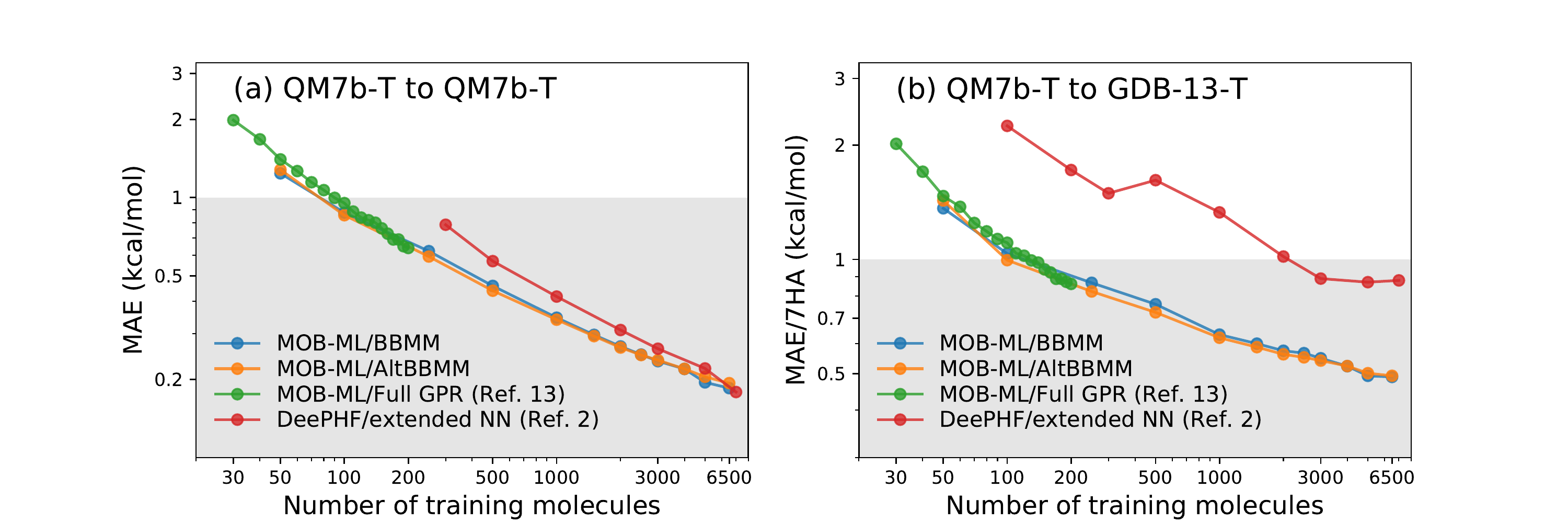}
	\caption{Learning curves for MOB-ML with different training protocols trained on QM7b-T and tested on (a) QM7b-T and (b) GDB-13-T. The accuracies of QM7b-T and GDB-13-T are measured by the MAEs and MAEs per 7 heavy atoms (MAE/7HA) of test molecules, respectively.
	We additionally plot the current best results in low and big data regimes, i.e., MOB-ML training with full GPR from Ref.~\citenum{husch2021improved} and the state-of-art DeePHF/extended NN from Ref.~\citenum{chen2020ground}, respectively.
	The gray shaded area represents the chemical accuracy of 1 kcal/mol.
	}
	\label{figure:lc}
\end{figure}

\subsection{BBMM and AltBBMM for energies of organic molecules}
We now examine the accuracy and transferability of BBMM and AltBBMM in learning QM7b-T and GDB-13-T molecular energies.
The transferability of MOB-ML is assessed by the MAEs per 7 heavy atoms (MAE/7HA) of test GDB-13-T molecules predicted by the models trained on QM7b-T molecules.
Table \ref{table:6500} lists the wall-clock time of training on 6500 QM7b-T molecules by BBMM and AltBBMM and the corresponding prediction MAEs on test QM7b-T and GDB-13-T molecules.
Similar to the results in Table \ref{table:noise}, by utilizing our AltBBMM approach, we gain a four-fold speedup in the training timings while only introducing 4\% and  1\% additional MAE in the prediction of QM7b-T and GDB-13-T, respectively, compared with BBMM.

In addition, we compare the performance of BBMM and AltBBMM with the results of the current most accurate literature methods, i.e., MOB-ML with full GP (MOB-ML/Full GP)~\cite{husch2021improved} and DeePHF with extended neural network regressor (DeePHF/extended NN)~\cite{chen2020ground}.
The literature results of MOB-ML/Full GP are only available with up to 220 training molecules due to the limited memory resources.
The introduction of BBMM and AltBBMM allows MOB-ML to scale up the training to 6500 molecules (over 1 million training pair energies) while retaining the accuracy and transferability compared with MOB-ML/full GP in Figure \ref{figure:lc}. By training on 6500 molecules, BBMM and AltBBMM reach the current best MAE/7HA for GDB-13-T as 0.490 kcal/mol and 0.493 kcal/mol, respectively.
In all the cases we tested, BBMM and AltBBMM provide a better accuracy on QM7b-T and a better transferability on GDB-13-T than DeePHF/extended NN.

\section{Conclusion}
In this work, we successfully apply the BBMM algorithm and a new alternative implementation, AltBBMM, to Gaussian process-training for the MOB-ML method on over a million pair energies.
Even though the use of BBMM alone increases our previously attainable training-set size limit over 30 times, our newly introduced AltBBMM implementation improves this further by offering a four-fold speed-up while maintaining high accuracy.
With the BBMM and AltBBMM approaches, MOB-ML models can be trained using datasets with over 6500 QM7b-T molecules, yielding the best accuracy to date for the QM7b-T and GDB-13-T datasets.
Future work will include  extension of the approach for the prediction of other molecular properties within the MOB-ML framework.

\begin{ack}
We thank Dr. J. Emiliano Deustua for helpful discussions. This work is supported in part by the U.S. Army Research Laboratory (W911NF-12-2-0023), the U.S. Department of Energy (DE-SC0019390), the Caltech DeLogi Fund, and the Camille and Henry Dreyfus Foundation (Award ML-20-196).   Computational resources were provided by the National Energy Research Scientific Computing Center (NERSC), a DOE Office of Science User Facility supported by the DOE Office of Science under contract DE-AC02-05CH11231.
\end{ack}

\medskip
{
\small
\bibliography{mobml, gp}
}

\end{document}